\input{aipcheck}

\documentclass[
    ,final            
  ]
  {aipproc}
 \usepackage{amsmath,amssymb}
 
\layoutstyle{6x9}

\begin{document}

\title{Analyses of third order Bose-Einstein correlation by means of Coulomb wave function}

\classification{25.75.-q, 25.75.Gz}
\keywords      {Bose-Einstein correlation, Coulomb wave function, three-body}

\author{Minoru Biyajima}{ address={Department of Physics, Shinshu University, Matsumoto, 390-8621, Japan} }
\author{Takuya Mizoguchi}{ address={Toba National College of Maritime Technology, Toba 517-8501, Japan} }
\author{Naomichi Suzuki}{ address={Department of Comprehensive Management, Matsumoto University, 
Matsumoto 390-1295, Japan} }

\begin{abstract}
In order to include a correction by the Coulomb interaction in Bose-Einstein correlations (BEC), the wave function for the Coulomb  scattering were introduced in the quantum optical approach to BEC in the previous work.  If we formulate the amplitude written by Coulomb wave functions according to the diagram for BEC in the plane wave formulation,  the formula for $3\pi^-$BEC becomes simpler than that of our previous work. We re-analyze the raw data of  $3\pi^-$BEC by NA44 and STAR Collaborations by this formula. Results are compared with the previous ones.
\end{abstract}

\maketitle

%
\section{Introduction}

Recently, in addition to the data on the charged two-body Bose-Einstein correlationiBEC), 
data on the three-body charged BEC have been reported~\cite{na44,phen02,wa88,star03}. 
In some papers~\cite{phen02,wa88}, Coulomb correction is done with fixed interaction region, 5 fm.
On the other hand, raw data with acceptance correction are reported~\cite{na44,star03}.  
In Ref.~\cite{alt99}, the $3\pi^-$BEC is calculated by the use of the Coulomb wave function with fixed source radius $R$.
We have analyzed the BEC without assuming the size of source radius using the Coulomb wave function~\cite{mizo01,biya04}.
In this paper, we would like to refine the formula for the $3\pi^-$BEC using the diagrammatic representation for BEC in the quantum optical (QO) approach~\cite{biya90,suzu92,suzu97}.

In order to describe the two-body charged BEC ( for example, $2\pi^-$system ), we should solve the Shr\"{o}dinger equation of Coulomb scattering. The regular solution at the origin of the Coulomb potential is given by,
 \begin{eqnarray}
   \psi_{{{\bf k}_i {\bf k}_j}}^C({\bf x}_i, {\bf x}_j) = \Gamma(1 + i\eta_{ij})
   e^{-\pi \eta_{ij}/2} e^{ i{\bf k}_{ij} \cdot {{\bf r}}_{ij} }
   F[-i\eta_{ij}, 1; ( k_{ij} r_{ij} - {\bf k}_{ij} \cdot {{\bf r}}_{ij} )],    \label{eq.int1}
 \end{eqnarray}
where, ${\bf r}_{ij} = {\bf x}_i - {\bf x}_j$, ${\bf k}_{ij} = ( m_j{\bf k}_i - m_i{\bf k}_j)/(m_i+m_j)$, 
$r_{ij} = |{{\bf r}}_{ij}|$, $k_{ij} = |{\bf k}_{ij}|$ and $\eta_{ij} = e_ie_j\mu_{ij}/k_{ij}$. $m_{ij}$ 
is the reduced mass of $m_i$ and $m_j$, $i, j=1, 2, 3$ and $i\neq j$.
 $F[a, b; x]$ is the confluent hypergeometric function, and $\Gamma(x)$ is the Gamma function.

The wave function of identical bose particles should be symmetrized. For two particle momentum density we have,
%
\begin{eqnarray*}
  N^{(2\pi^-)}  &=&  \frac{1}{2} \prod_{i=1}^2 \int \rho({\bf x}_i) d^3 {\bf x}_i  
     \bigl| \psi_{{\bf k}_1{\bf k}_2}^C({\bf x}_1,\ {\bf x}_2) 
      + \psi_{{\bf k}_1{\bf k}_2}^C({\bf x}_2,\ {\bf x}_1) \bigr|^2   \\ \nonumber
               &=& \prod_{i=1}^2 \int \rho({\bf x}_i) d^3 {\bf x}_i  (G_1 + G_2 ), \\
   G_1 &=& \frac{1}{2} \Bigl( \bigl| \psi_{{\bf k}_1{\bf k}_2}^C({\bf x}_1,\ {\bf x}_2) \bigr|^2
      + \bigl|\psi_{{\bf k}_1{\bf k}_2}^C({\bf x}_2,\ {\bf x}_1) \bigr|^2  \Bigr),   \nonumber \\
   G_2 &=& {\rm Re}\Bigl(  \psi_{{\bf k}_1{\bf k}_2}^C({\bf x}_1,\ {\bf x}_2) 
      \psi_{{\bf k}_1{\bf k}_2}^{C*}({\bf x}_2,\ {\bf x}_1)  \Bigr),       
      \label{eq.int2} 
\end{eqnarray*}
where,
 \begin{eqnarray}
     \rho ({\bf x})=\frac{1}{({2\pi R^2})^{3/2}}\exp[-\frac{{\bf x}^2}{2R^2}]. \label{eq.int3}
 \end{eqnarray}
The three particle momentum density for 3$\pi^-$ BEC is written as~\cite{alt99},
 \begin{eqnarray}
  N^{(3\pi^-)} &=& \frac{1}{6} \prod_{i=1}^3 \int \rho({\bf x}_i) d^3 {\bf x}_i 
    \bigl| \sum_{j=1}^6 A(j) \bigr|^2,  \label{eq.int4a} \\
  A(1) &=& A_1= 
  \psi_{{\bf k}_1{\bf k}_2}^C({\bf x}_1,\ {\bf x}_2)
  \psi_{{\bf k}_2{\bf k}_3}^C({\bf x}_2,\ {\bf x}_3)
  \psi_{{\bf k}_3{\bf k}_1}^C({\bf x}_3,\ {\bf x}_1), \nonumber\\
 A(2) &=& A_{23} =
     \psi_{{\bf k}_1{\bf k}_2}^C({\bf x}_1,\ {\bf x}_3) 
     \psi_{{\bf k}_2{\bf k}_3}^C({\bf x}_3,\ {\bf x}_2) 
     \psi_{{\bf k}_3{\bf k}_1}^C({\bf x}_2,\ {\bf x}_1), \nonumber \\
  A(3) &=& A_{12}=
      \psi_{{\bf k}_1{\bf k}_2}^C({\bf x}_2,\ {\bf x}_1)
      \psi_{{\bf k}_2{\bf k}_3}^C({\bf x}_1,\ {\bf x}_3)
      \psi_{{\bf k}_3{\bf k}_1}^C({\bf x}_3,\ {\bf x}_2), \nonumber\\
  A(4)&=& A_{123}=
      \psi_{{\bf k}_1{\bf k}_2}^C({\bf x}_2,\ {\bf x}_3) 
      \psi_{{\bf k}_2{\bf k}_3}^C({\bf x}_3,\ {\bf x}_1) 
      \psi_{{\bf k}_3{\bf k}_1}^C({\bf x}_1,\ {\bf x}_2), \nonumber\\
  A(5) &=& A_{132} = 
     \psi_{{\bf k}_1{\bf k}_2}^C({\bf x}_3,\ {\bf x}_1)
     \psi_{{\bf k}_2{\bf k}_3}^C({\bf x}_1,\ {\bf x}_2)
     \psi_{{\bf k}_3{\bf k}_1}^C({\bf x}_2,\ {\bf x}_3), \nonumber\\
  A(6) &=& A_{13}=
        \psi_{{\bf k}_1{\bf k}_2}^C({\bf x}_3,\ {\bf x}_2) 
        \psi_{{\bf k}_2{\bf k}_3}^C({\bf x}_2,\ {\bf x}_1) 
        \psi_{{\bf k}_3{\bf k}_1}^C({\bf x}_1,\ {\bf x}_3).  \label{eq.int4}
\end{eqnarray}
In the plane wave approximation, each amplitude $A(i)$ approaches to the following form;
\begin{eqnarray}
  \label{eq.qoa1}
  A(1) &=&  A_1 \stackrel{\rm {PW}}{\longrightarrow}
          e^{ i {\bf k}_{12} \cdot {\bf r}_{12}}
          e^{ i {\bf k}_{23} \cdot {\bf r}_{23}}
          e^{ i {\bf k}_{31} \cdot {\bf r}_{31}}
          = e^{ (3/2)i ({\bf k}_1 \cdot {\bf x}_1
          + {\bf k}_2 \cdot {\bf x}_2
          + {\bf k}_3 \cdot {\bf x}_3)}, \nonumber \\
 A(2)&=& A_{23} \stackrel{\rm {PW}}{\longrightarrow} 
         e^{ i {\bf k}_{12} \cdot {\bf r}_{13}} 
         e^{ i {\bf k}_{23} \cdot {\bf r}_{32}} 
         e^{ i {\bf k}_{31} \cdot {\bf r}_{21}}
         = e^{ (3/2)i ({\bf k}_1 \cdot {\bf x}_1
          + {\bf k}_2 \cdot {\bf x}_3
          + {\bf k}_3 \cdot {\bf x}_2)}, \nonumber \\
  A(3) &=&  A_{12} \stackrel{\rm {PW}}{\longrightarrow}
          e^{ i {\bf k}_{12} \cdot {\bf r}_{21}}
          e^{ i {\bf k}_{23} \cdot {\bf r}_{13}}
          e^{ i {\bf k}_{31} \cdot {\bf r}_{32}}
          = e^{ (3/2)i ({\bf k}_1 \cdot {\bf x}_2
          + {\bf k}_2 \cdot {\bf x}_1
          + {\bf k}_3 \cdot {\bf x}_3)}, \nonumber \\
  A(4)&=& A_{123} \stackrel{\rm {PW}}{\longrightarrow}
          e^{ i {\bf k}_{12} \cdot {\bf r}_{23}} 
          e^{ i {\bf k}_{23} \cdot {\bf r}_{31}}
          e^{ i {\bf k}_{31} \cdot {\bf r}_{12}}
          = e^{ (3/2)i ({\bf k}_1 \cdot {\bf x}_2
          + {\bf k}_2 \cdot {\bf x}_3
          + {\bf k}_3 \cdot {\bf x}_1)}, \nonumber \\
  A(5)&=& A_{132} \stackrel{\rm {PW}}{\longrightarrow}
          e^{ i {\bf k}_{12} \cdot {\bf r}_{31}}
          e^{ i {\bf k}_{23} \cdot {\bf r}_{12}}
          e^{ i {\bf k}_{31} \cdot {\bf r}_{23}}
          = e^{ (3/2)i ({\bf k}_1 \cdot {\bf x}_3
          + {\bf k}_2 \cdot {\bf x}_1
          + {\bf k}_3 \cdot {\bf x}_2)}, \nonumber \\
  A(6) &=& A_{13} \stackrel{\rm {PW}}{\longrightarrow} 
           e^{ i {\bf k}_{12} \cdot {\bf r}_{32}} 
           e^{ i {\bf k}_{23} \cdot {\bf r}_{21}} 
           e^{ i {\bf k}_{31} \cdot {\bf r}_{13}}
          = e^{ (3/2)i ({\bf k}_1 \cdot {\bf x}_3
          + {\bf k}_2 \cdot {\bf x}_2
          + {\bf k}_3 \cdot {\bf x}_1)},  \label{eq.int5}
\end{eqnarray}
where PW means the plane wave approximation of the amplitude.
It should be noted that factor 3/2 does not appear in the formulation of $3\pi^-$BEC using the plane wave.

\section{Quantum optical approach}

The amplitude squared in Eq.~(\ref{eq.int4a}) can be classified 
into the following groups,
\begin{eqnarray}
  F_1 &=& ({1}/{6}) [ A_{1}A^*_{1} + A_{12}A^*_{12} + A_{23}A^*_{23}
          + A_{13}A^*_{13} + A_{123}A^*_{123} + A_{132}A^*_{132} ],  \nonumber \\
  F_{12} &=& ({1}/{6}) [ A_{1}A^*_{12} + A_{23}A^*_{123} + A_{13}A^*_{132} + c.c.], \nonumber \\
  F_{23} &=& ({1}/{6}) [ A_{1}A^*_{23} + A_{12}A^*_{132} + A_{13}A^*_{123} + c.c.],  \nonumber \\
  F_{31} &=& ({1}/{6}) [ A_{1}A^*_{13} + A_{23}A^*_{132} + A_{12}A^*_{123} + c.c. ], \nonumber \\
  F_{123} &=& ({1}/{6}) [ A_{1}A^*_{132} + A_{132}A^*_{123} + A_{13}A^*_{12} 
                + A_{12}A^*_{23} + A_{23}A^*_{13} + A_{123}A^*_{1}]  \nonumber \\
  F_{132} &=& ({1}/{6}) [ A_{1}A^*_{123} + A_{23}A^*_{12} + A_{12}A^*_{13} 
                + A_{123}A^*_{132} + A_{132}A^*_{1} + A_{13}A^*_{23}],             \label{eq.qoa2} 
\end{eqnarray}
where, c.c. denotes the complex conjugate. In the plane wave approximation, $F_1$ reduces to 1, $F_{ij}$ corresponds to exchange between $i$ and $j$ charged particles, and $F_{123}$ correspond to exchange among three charged particles. 

In the previous work~\cite{mizo01}, we introduced the Coulomb wave function to the core-halo model~\cite{csor99}, where following parameters are used; the fraction of multiplicity from the core part, $f_c=\langle n_{core}\rangle/\langle n_{tot}\rangle$, and the fraction of coherently produced particles from the core part,  $p_c=\langle n_{co}\rangle/\langle n_{core}\rangle$.  In the quantum optical approach, chaoticity parameter is defined as $p=\langle n_{chao}\rangle/\langle n_{core}\rangle$, where $p=1-p_c$. In the following, we use $f_c$ and $p$. The radius of halo part is assumed to be infinitely large. Therefore, particles emitted from the halo part do not contribute to BEC, namely exchange of particles.  In Ref.~\cite{suzu97}, the higher order BEC is formulated, where a contamination effect is included. 
If we can identify the contribution of the halo part to the contamination, both formulations coincide.
%
 \begin{figure}[htb]
    \includegraphics[scale=0.40]{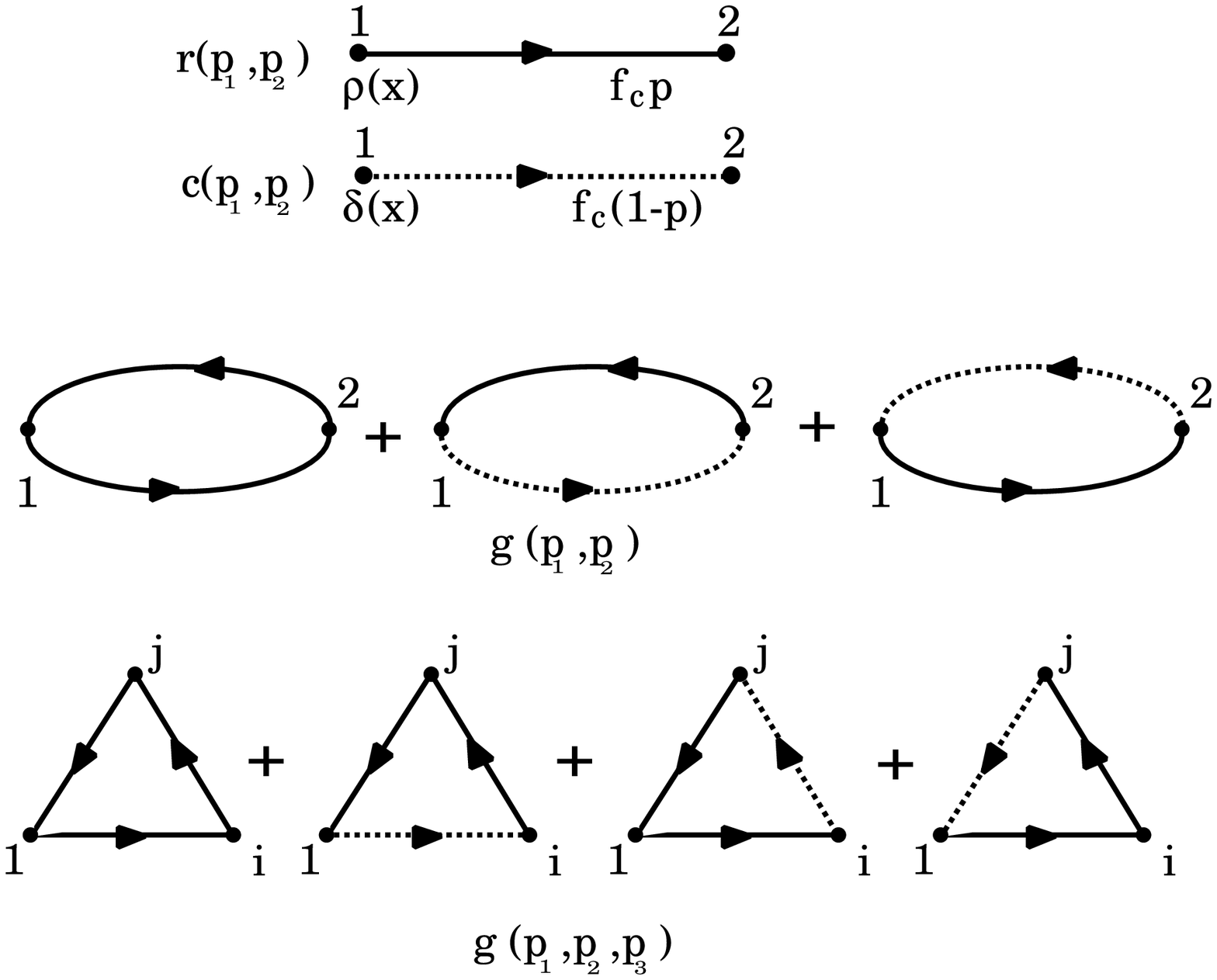}
  \caption{Cumulant up to the third order, $(i,j)=(2,3)$ or $(3,2)$.}
  \label{fig.cumu-3rd}
 \end{figure}
%
 
In the QO approach, the two particle momentum density $\rho({\bf p}_1,{\bf p}_2)$ and the three particle momentum density $\rho({\bf p}_1,{\bf p}_2,{\bf p}_3)$  are given respectively as~\cite{suzu92,suzu97},
\begin{eqnarray*}
  \rho({\bf p}_1,{\bf p}_2)&=&\rho({\bf p}_1)\rho({\bf p}_2)+g({\bf p}_1,{\bf p}_2),  \\
  \rho({\bf p}_1,{\bf p}_2,{\bf p}_3)&=&\rho({\bf p}_1)\rho({\bf p}_2)\rho({\bf p}_3)
         +g({\bf p}_1,{\bf p}_2)\rho({\bf p}_3) \\
  &&+g({\bf p}_2,{\bf p}_3)\rho({\bf p}_1)+g({\bf p}_3,{\bf p}_1)\rho({\bf p}_2) + g({\bf p}_1,{\bf p}_2,{\bf p}_3).
\end{eqnarray*}
The second order cumulant $g({\bf p}_1,{\bf p}_2)$ and the third order cumulant 
$g({\bf p}_1,{\bf p}_2,{\bf p}_3)$ are given in Fig.~\ref{fig.cumu-3rd}~\cite{suzu92,suzu97}.  
In  Fig.~\ref{fig.cumu-3rd}, the chaotic component $r({\bf p}_1,{\bf p}_2)$ is shown by 
the solid line with arrow from 1 to 2, and the coherent component 
$c({\bf p}_1,{\bf p}_2)$ is shown by dotted line. 
In the formulation of BEC with the Coulomb wave function, 
momentum and coordinate cannot be decoupled. Therefore, 
interpretation of diagram is somewhat modified; 
 The source function $\rho({\bf x})$ is attached to the starting point of solid line, 
 and the delta function $\delta({\bf x})$ is to that of dashed line.
In addition, $p$ is replaced by $f_cp$, and $1-p$ is by $f_c(1-p)$ to include the contribution 
from the halo part.

In the $2\pi^-$BEC, $N^{2\pi^-}$ corresponds to $\rho({\bf p}_1,{\bf p}_2)$, 
$G_1$ to $\rho({\bf p}_1)\rho({\bf p}_2)$, 
and $G_2$ to $g({\bf p}_1,{\bf p}_2)$ beside the normalization factor.
Then, the formula for $2\pi^-$BEC is written as,   
 \begin{eqnarray*}
   \frac{ N^{2\pi^-} }{ N^{BG} } &=& 
        C\prod_{i=1}^2 \int \rho({\bf x}_i) d^3 {\bf x}_i  (G_1 + f_c^2 p^2G_2 ) \\
    && +C\int d^3{\bf x}_1\int d^3{\bf x}_2 \rho({\bf x}_1)\delta({\bf x}_2) 2f_c^2p(1-p)G_2 
%
 %
 \end{eqnarray*}

In the third order BEC, $N^{3\pi^-}$ corresponds to $\rho({\bf p}_1,{\bf p}_2,{\bf p}_3)$, 
$F_1$ to $\rho({\bf p}_1)\rho({\bf p}_2)\rho({\bf p}_3)$,  $F_{ij}$ to $g({\bf p}_i,{\bf p}_j)$, 
and ($F_{123} + F_{123}$) to $g({\bf p}_1,{\bf p}_2,{\bf p}_3)$, beside the normalization factor.
Then, the formula for $3\pi^-$BEC is given by
\begin{eqnarray}
     \frac{N^{3\pi^-}}{N^{BG}} &=& C \prod_{i=1}^3  \int \rho({\bf x}_i) d^3 {\bf x}_i
            \left[ F_1 + 3f_c^2p^2F_{12} + 2f_c^3p^3\cdot Re[F_{123}]  \right]  \nonumber  \\
  &+& C \prod_{i=1}^3 \int  d^3 {\bf x}_i \delta({\bf x}_1) \rho({\bf x}_2) \rho({\bf x}_3)  
        6 f_c^2p(1-p) \bigl( F_{12} + f_cp Re[F_{123}] \bigr).      \label{eq.qoa4}
\end{eqnarray}
Compare Eq.(\ref{eq.qoa4}) with Eq.(12) in Ref.~\cite{biya04}, where $F_2=F_{12}+F_{23}+F_{31}$  and $F_3=F_{123}+F_{132}$.
%
 %
%

\section{Analysis of $3\pi^-$BEC}

The formula (\ref{eq.qoa4}) is applied to the analysis of raw data on $3\pi^-$BEC by STAR Collaboration~\cite{star03} and NA44 Collaboration~\cite{na44}.
%
\begin{table}[htb]
 \begin{tabular}{llcccc} 
   \hline
       & $p$              &   1.0         &   0.8         &   0.6          &  0.40$\pm$0.02 \\    \hline
 present work  &     $f_c$            & 0.75$\pm$0.02 & 0.78$\pm$0.02 & 0.85$\pm$0.02  & 1.0(fixed) \\    
    &    $R$ (fm)         & 5.34$\pm$0.24 & 6.21$\pm$0.30 & 6.88$\pm$0.35  & 7.43$\pm$0.34 \\   
   & $\chi^2/N_{dof}$     &   2.68/34     &   0.93/34     &    0.65/34     & 0.58/34 \\ 
%
%
  \hline
 Ref.~\cite{mizo01} &    $f_c$          & 0.75$\pm$0.02 & 0.78$\pm$0.02 & 0.85$\pm$0.02 & \\    
     &  $R$ (fm)          & 5.34$\pm$0.24 & 5.99$\pm$0.28 & 6.48$\pm$0.32 & \\   
   & $\chi^2/N_{dof}$     &   2.80/34     &   1.39/34     & 0.98/34       & \\ \hline 
  \end{tabular}
  \caption{Analyses of raw data on $3\pi^-$ BEC by STAR Collaboration~\cite{star03}. }\\
     \label{table1}
\end{table}
%

\begin{figure}[h]
 \begin{minipage}{0.45\linewidth}
    \includegraphics[scale=0.80]{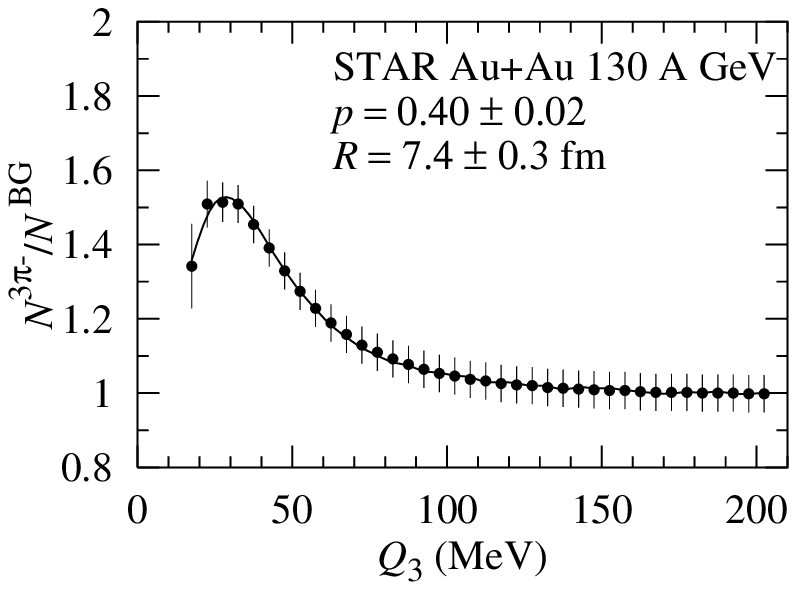}
 \end{minipage}
 \begin{minipage}{0.45\linewidth}
    \includegraphics[scale=0.38]{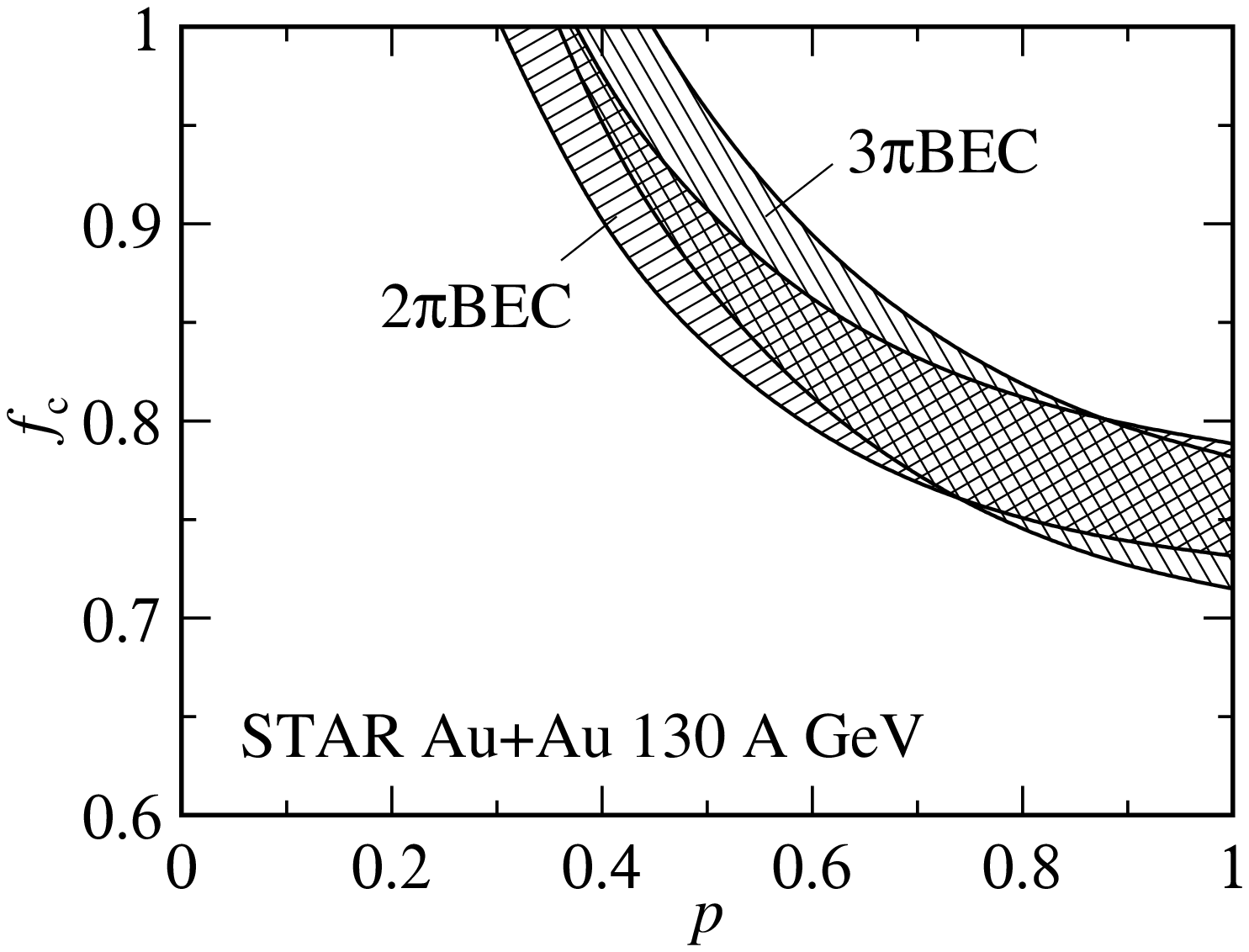}
    \caption{(a) Analysis of raw data on $3\pi^-$BEC by STAR Collaboration; 
    $f_c$ is fixed at 1.  (b) Possible region of $f_c$ and $p$ estimated from the analyses of 
    $2\pi^-$ BEC and $3\pi^-$BEC.}
 \label{fig.star}
 \end{minipage}
\end{figure}

The results for $3\pi^-$BEC by STAR Collaboration~\cite{star03} are shown in Table~\ref{table1} and Fig.~\ref{fig.star}. For comparison, the results in the previous work~\cite{mizo01} are also shown in the lower part of Table~\ref{table1}.
Parameters $f_c$ and $p$ are almost the same with the previous calculations. 
The radius $R$ is almost the same at $p=1.0$, and the present result gradually becomes larger than the previous one. 
 At $p=0.6$, the new result becomes about 6\% larger than the previous one .
Fitting to the STAR data is slightly improved. 
In Fig.\ref{fig.star}, the result at p=1 is shown, and the possible region of $p$ and $f_c$ estimated from the analysis of $2\pi^-$BEC and $3\pi^-$BEC is shown. Each hatched region is defined by the fitted value $\pm$ 2 standard deviation.
%
%
%
%
\begin{table}[htb]
 \begin{tabular}{llcccc}
  \hline
  &   $p$              & 1.0       & 0.8        & 0.6         &  0.32$\pm$0.05 \\   \hline
 present work   & $f_c$       & 0.67$\pm$0.04 & 0.70$\pm$0.04 & 0.76$\pm$0.05 & 1.0(fixed) \\    
    & $R$ (fm)             & 2.88$\pm$0.39 & 3.32$\pm$0.49 & 3.66$\pm$0.54 & 4.05$\pm$0.57 \\   
   &  $\chi^2/N_{dof}$     &  6.64/15      & 6.78/15       & 6.81/15       & 6.80/15 \\ 
%
%
  \hline
 Ref.~\cite{biya04}   & $f_c$                & 0.67$\pm$0.04 & 0.70$\pm$0.04 & 0.76$\pm$0.05 &  \\    
    & $R$ (fm)             & 2.89$\pm$0.39 & 3.22$\pm$0.46 & 3.47$\pm$0.51 & \\   
    & $\chi^2/N_{dof}$     &  6.7/15       & 6.8/15        & 6.8/15        & \\ \hline 
  \end{tabular}  
   \caption{Analyses of raw data on $3\pi^-$ BEC by NA44 Collaboration~\cite{na44}. }\\
    \label{table2}
\end{table} 
%
\begin{figure}[h]
 \begin{minipage}{0.45\linewidth}
    \includegraphics[scale=0.8]{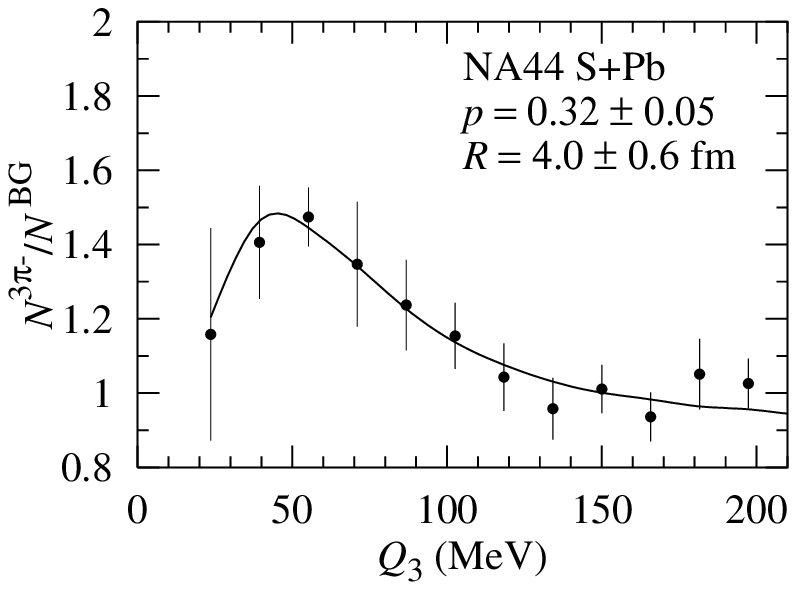}
 \end{minipage}
 \begin{minipage}{0.45\linewidth}
    \includegraphics[scale=0.38]{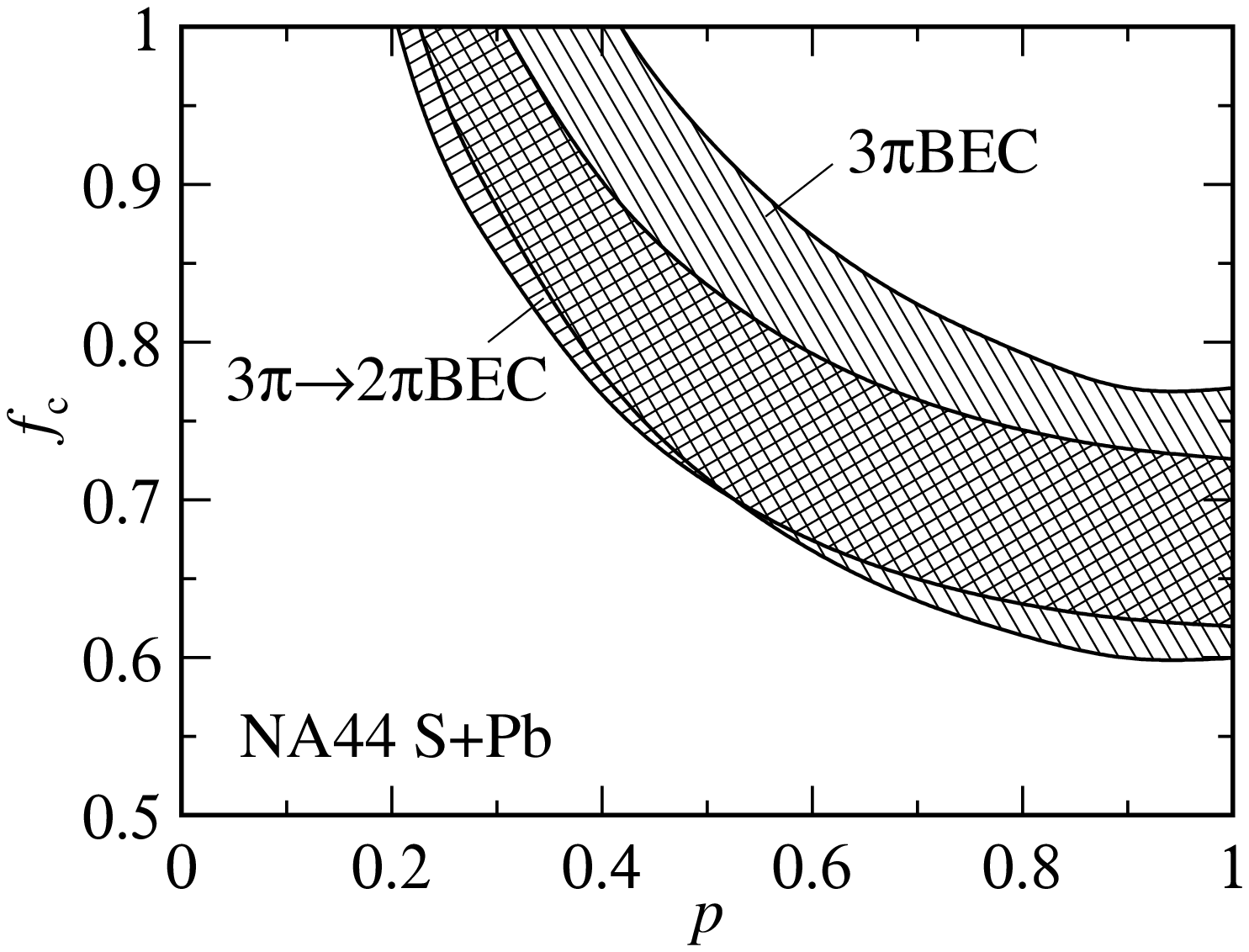}
    \caption{(a) Analysis of $3\pi^-$BEC by NA44 Collaboration; $f_c$ is fixed at 1.  
    (b) Possible region of $f_c$ and $p$ estimated from the analyses of $2\pi^-$ BEC and $3\pi^-$BEC. }
 \label{fig.na44}
 \end{minipage}
\end{figure}
%

 The results for $3\pi^-$BEC by NA44 Collaboration~\cite{na44} are shown in Table~\ref{table2} and Fig.~\ref{fig.na44}. Previous results~\cite{biya04} are also shown in Table~\ref{table2}.
 Parameters $f_c$ and $p$ are almost the same with the previous calculations. 
As for the radius $R$, the present result gradually becomes larger than the previous one, as the chaoticity parameter decreases from 1 to 0.6. 
 At $p=0.6$, the new result becomes about 5\% larger than the previous one.
Fitting of the present calculations to the NA44 data is almost the same with that of the previous results.

\section{Concluding remarks}

We refine the formula for $3\pi^-$BEC according to the diagram in the QO approach.
In our formulation three parameters, fraction of core part $f_c$, chaoticity parameter $p$, 
and radius of interaction range $R$ are included.
We apply the formula to the analysis of data on $3\pi^-$BEC by STAR and NA44 Collaborations, 
and compare with the previous calculations.
Parameters $f_c$ and $p$ are almost the same with the previous calculations. 
The radius $R$ is almost the same at $p=1.0$, and the present result gradually becomes larger than the previous one, as $p$ decreases. 
 At $p=0.6$, the new result becomes about 5\% larger than the previous one.
Fitting to the STAR data is slightly improved. That to the NA44 data is almost the same.

From the comparison of the results on $2\pi^-$BEC~\cite{mizo01,biya04} with those on $3\pi^-$BEC, 
we have a relation,
\begin{eqnarray}
   R_{2nd}\simeq 1.5R_{3rd}. \label{eq.fin1}
\end{eqnarray}
Factor 1.5 in Eq.~(\ref{eq.fin1}) corresponds to $3/2$ in Eq.~(\ref{eq.int5}).
This result suggests that our approach should be equivalent to the plane wave formulation applied 
to the data with Coulomb correction.  Detail calculation is reported in Ref.~\cite{suzu05f}

%
%

%

\begin{theacknowledgments}
  Authors would like to thank J.R.Glauber for variable comments.
  They also would like to thank RCNP at Osaka University, Faculty of Science, 
  Shinshu University,  Toba National College of Maritime Technology 
  and Matsumoto University for financial support.
\end{theacknowledgments}


\end{document}